\documentclass[final,3p,times,twocolumn]{elsarticle}

\usepackage{amssymb}
\usepackage{amsmath,amssymb,exscale}
\usepackage{graphicx}
\usepackage{epsfig}
\usepackage{multicol}
\usepackage{color}
\usepackage{xcolor}
\usepackage{hyperref}
\usepackage{mathrsfs}
\usepackage{hyperref}
\hypersetup{colorlinks,bookmarksopen,bookmarksnumbered,citecolor=rossos,
linkcolor=black,pdfstartview=FitH,urlcolor=blus}
\usepackage{amsfonts}
\usepackage{slashed}
\usepackage{blindtext}
 \usepackage{fancyhdr}
\usepackage{hyperref}
\usepackage{mathtools}
\biboptions{numbers,sort&compress}

\allowdisplaybreaks

\definecolor{blus}{cmyk}{1,1,0,0.}
\definecolor{verdes}{cmyk}{0.99,0,0.99,0.02}
\definecolor{rossos}{cmyk}{0,1,1,0.55}
\definecolor{greeny}{cmyk}{0.99,0,0.59,0.98}
\definecolor{redy}{cmyk}{0,1,1,0.40}

\newcommand{\bp}{M_P}

\def\be{\begin{equation}}
\def\ee{\end{equation}}
\def\bea{\begin{eqnarray}}
\def\eea{\end{eqnarray}}

\def\ba{\begin{array} }
\def\ea{\end{array}}
\def\bac{\begin{array} {c}}
\def\bacc{\begin{array} {cc}}
\def\baccc{\begin{array} {ccc}}

\def\psl{\hbox{\hbox{${p}$}}\kern-1.9mm{\hbox{${/}$}}}
\def\dsl{\hbox{\hbox{${\partial}$}}\kern-1.7mm{\hbox{${/}$}}}
\def\Dsl{\hbox{\hbox{${D}$}}\kern-2.1mm{\hbox{${/}$}}}

\definecolor{red}{rgb}{1,0,0}

\def\hhref#1{\href{http://arxiv.org/abs/#1}{arXiv:#1}}  
\hypersetup{colorlinks=true, linkcolor=red}
 
\journal{the arXiv}

\begin{document}

\begin{frontmatter}

\title{\vspace{-2cm}\huge {\color{redy} Unimodular Quadratic Gravity \\ and the Cosmological Constant
}}  

\author{\vspace{1cm}{\large {\bf Alberto Salvio}}}

\address{\normalsize \vspace{0.2cm}Physics Department, University of Rome and INFN Tor Vergata, Italy \\

\vspace{0.3cm} 
 \vspace{-1cm}
 }

\begin{abstract}
\noindent Unimodular gravity addresses the {\it old} cosmological constant (CC) problem, explaining why such constant is not at least as large as the largest particle mass scale, but classically it is indistinguishable from ordinary gravity. 
 Conversely, quantum physics may give us a  way to distinguish the two theories. Thus, here the unimodular constraint is imposed on a  non-perturbative and background-independent quantum version of quadratic gravity, which was recently formulated. It is shown that unimodularity does lead to different predictions for some  inflationary quantum observables. 
Unimodular gravity {\it per se} does not solves the new CC problem (why the CC has the observed value?) even in this realization. To address this issue a multiverse made by different eras in a single big bang is considered and the observed scale of dark energy is explained anthropically.

\end{abstract}


\end{frontmatter}
 
 \begingroup
\hypersetup{linkcolor=blus}
\tableofcontents
\endgroup


\section{Introduction}\label{introduction}
  
  There is more than one issue related to the observed value of the CC. The old CC problem consists in explaining why the CC is not at least of the same order of magnitude of the largest particle mass scale. This is because each particle contributes to the vacuum energy density
  through a term of order of its mass to the fourth and in Einstein gravity vacuum energy density contributes to the CC. The new CC problem consists in understanding why it is comparable to the present matter density~\cite{Weinberg:2000yb} although it scales differently with time; this issue is also known as the coincidence problem. Both problems are only fine-tunings, they do not indicate inconsistencies between theory and observations. However, solutions may suggest routes to search for new physics.
  
  In some theories the CC is promoted to a dynamical scalar field with a potential that is so slowly varying to mimic a CC. Even in these realizations the CC problems persist: such potential needs to be fine-tuned because each particle still contributes to its zero-point value as described above, and no explanation between the current comparable values of dark energy and matter densities is provided.  The Euclid satellite~\cite{Euclid}, which was launched on July 1, 2023, will soon provide information on the nature of dark energy and further increase the interest in this field of fundamental physics. 
  
  In unimodular gravity (UG) one requires by definition that the spacetime volume is not a dynamical degree of freedom (see~\cite{Weinberg:1988cp} and references therein). This constraint effectively changes the nature of the CC from the coefficient of a term in the action (which the vacuum energy density contributes to) to an integration constant of the classical field equations, regardless of the theory on which this constraint is imposed~\cite{Percacci:2017fsy}. Therefore, in the presence of the unimodular constraint there is no reason to expect that the CC is at least of the same order of magnitude of the largest particle mass scale, because vacuum energy no longer gravitates. Still, the new CC problem remains unsolved because UG does not suggest any preferred value for this constant. Anthropic considerations~\cite{Weinberg:1988cp,Weinberg:1987dv} may explain the value of the CC, but require a multiverse, which so far has led to complicated  landscapes, where theoretical control is typically lost. 

  The main purpose of this paper is to combine UG and the anthropic principle to address the CC problems. 
    
  Given the relevance of UG, it is also important to look for observational tests. Classically, the unimodular constraint does not change the field equations, but only the theoretical nature of the CC; as a result, UG turns out to predict just the same physics as gravity without the unimodular constraint. While this is reassuring, because it implies that UG is a viable modification of gravity, it is also disappointing because classical physics does not allow us to distinguish between the two theories.
  
  Quantum mechanics, however, can change the situation completely as there is no theorem establishing the physical equivalence at quantum level. In order to understand if this really happens a consistent quantum gravity theory must be considered. In this paper we implement the unimodular constraint in quadratic gravity, a renormalizable~\cite{Weinberg:1974tw,Deser:1975nv,Stelle:1976gc,Barvinsky:2017zlx} and unitary~\cite{Salvio:2019wcp,Salvio:2020axm} UV extension of Einstein gravity, which was recently  formulated in a  non-perturbative and background-independent way\footnote{The Higgs mass fine-tuning problem in quadratic gravity has been previously addressed in~\cite{Salvio:2014soa,Kannike:2015apa,Salvio:2017qkx}.}~\cite{Salvio:2024joi}. 
  
  The classical action of quadratic gravity we consider is
\be  S^{\rm ren} = \int d^4x \sqrt{-g}\left( \frac{R^2}{6f_0^2}-\frac{1}{2f_2^2} W^2+\frac{\bp^2}{2} R - \Lambda_0\right)\label{TotAction}. \ee
Here $f_0, f_2, \bp$ and $\Lambda_0$ are renormalized parameters, $g$ is the determinant of the metric, $R$ is the Ricci scalar and $W^2\equiv W_{\mu\nu\rho\sigma}W^{\mu\nu\rho\sigma}$ is the ``square" of the Weyl tensor $W_{\mu\nu\rho\sigma}$. 

We will show that a natural implementation of the unimodular constraint in quantum quadratic gravity  is possible through the path integral formalism. This is the first time a  non-perturbative and background-independent quantum UG is formulated.

Theoretical differences between quadratic gravity and its unimodular counterpart must be present at quantum level because, as we will show explicitly, in the latter, unlike in the former, one path integrates only over those metrics respecting the unimodular constraint. Theoretical differences in the context of Einstein gravity have been noted in~\cite{Eichhorn:2013xr,Bufalo:2015wda}. However, observational differences are necessary to physically distinguish the two theories.

The natural arena to look for quantum gravity observables is inflation. In this paper we will then focus on that period of the cosmological expansion and find indeed observational differences.

Here, post-inflationary physics is also considered, where a multiverse made by different eras in a single big bang is present, to find an explanation of the observed CC scale (the new CC problem). This leads to a landscape of values of the CC that are scanned during different eras. Such landscape, however, does not need to be complex as it would have to without the UG solution to the old CC problem.

  \vspace{-0.3cm}

  \section{Non-perturbative quantum quadratic gravity}\label{Non-perturbative quantum quadratic gravity}
  \vspace{-0.2cm}

In the non-perturbative and background-independent formulation of quantum quadratic gravity of~\cite{Salvio:2024joi}, the canonical coordinates $q$ are initially identified in the Gauss spacetime coordinate system and are the values of the 3D metric $g_{ij}$ and its time derivative, $K_{ij} \equiv- \dot g_{ij}/2$. Let us start by reviewing the findings of~\cite{Salvio:2024joi}, which are necessary to understand the original results of this paper. In the next section a unimodular version will be constructed. While $g_{ij}$ and its conjugate momentum are quantized in the ordinary way, $K_{ij}$ and its conjugate momentum are subject to an alternative quantization first discussed by Pauli~\cite{Pauli}, who elaborated on a previous work by Dirac~\cite{Dirac}. This Dirac-Pauli (DP) quantization has been more recently developed in~\cite{Salvio:2015gsi} (see also~\cite{Salvio:2018crh,Salvio:2020axm,Salvio:2024joi} for reviews).
 The Euclidean path integral for the transition amplitudes (between states of definite canonical coordinates $g_{ij}$ and $K_{ij}$) in the presence of an external  ``current" $J^{ij}$ for $g_{ij}$ is 
\bea &&\langle q_{f\eta},\tau_f|q_i,\tau_i\rangle^J  = \nonumber \\ \hspace{-1.9cm}&& \hspace{-.9cm}\int^{q(\tau_f) = q_f}_{q(\tau_i) = q_i}  \, C\delta g \, \exp\left(-S_E  + \int_{\tau_i}^{\tau_f} d\tau\int d^3x \, J^{ij}g_{ij}\right), \label{PIEucQG}\eea
where $S_E$ is the Euclidean action of quadratic gravity with bare parameters.
The boundary conditions at initial and final imaginary times, $\tau_i$ and $\tau_f$, respectively, are
\bea && q(\tau_i) = q_i: \quad g_{lm}(\tau_i) = g^{(i)}_{lm},~g'_{lm}(\tau_i) = -2K_{lm}^{(i)}, \label{gKboundp}\\ && q(\tau_f) = q_f: \quad g_{lm}(\tau_f) = g^{(f)}_{lm},~g'_{lm}(\tau_f) = -2K_{lm}^{(f)}, \label{gKbound} \eea
where $g^{(i,f)}_{lm}$ and $K^{(i,f)}_{lm}$ provides initial and final conditions for the metric and its time derivative, a prime denotes a derivative with respect to the imaginary time $\tau$ and, for simplicity, the dependence on the spatial coordinates is understood in~(\ref{gKboundp}) and~(\ref{gKbound}).  Also a label $\eta$ indicates the sign reversal of the canonical variables that are DP quantized; this ensures that the corresponding inner product is positive-definite. The integration measure $C \delta g$ over the 3D metrics is invariant under 3D general coordinate transformations.

In a generic spacetime coordinate system, on the other hand,
\bea &&\langle q_{f\eta},\tau_f|q_i,\tau_i\rangle^J  = \int^{q(\tau_f) = q_f}_{q(\tau_i) = q_i}  \, {\cal D}g \,  \left|\det\frac{\partial f}{\partial\xi}\right|\, \delta(f)  \nonumber \\&&\times  \exp\left(-S_E  + \int_{\tau_i}^{\tau_f} d\tau\int d^3x J^{\mu\nu}g_{\mu\nu}\right), \label{PIEucQGFP}\eea
where the metric measure ${\cal D}g$ is invariant under 4D general coordinate transformations (greek letters denote 4D spacetime coordinates), the four spacetime functions $\xi$ correspond to the 4D diffeomorphisms  and $f$ plays the role of a gauge-fixing function (its choice corresponds to the choice of the coordinate system).

Through~(\ref{PIEucQGFP}) one can also obtain the generating functional of Green's function, which reads (choosing this time the Lorentzian signature)
\bea &&{\cal Z}(J) = \frac1{``J\to 0"}\int  \, {\cal D}g \,  \left(\det\frac{\delta f}{\delta\xi}\right)\, \delta(f) \nonumber \\ &&\times \exp\left(iS  + i\int d^4x \,J^{\mu\nu}g_{\mu\nu}\right), \label{PIEucGF2L}\eea
where $S$ is the classical Lorentzian action with bare parameters and  the denominator $``J\to 0"$ recalls us that the path integral as usual should be divided  by the same quantity but with vanishing external 4D ``current", $J^{\mu\nu}=0$

  \vspace{-0.1cm}
  
 \section{General unimodular constraint}\label{General unimodular constraint}

In unimodular gravities (including quadratic gravity) one requires that the volume of spacetime is not a dynamical variable, but rather a fixed quantity. Mathematically, this constraint can be imposed by inserting in the Euclidean path integral~(\ref{PIEucQG}) the (functional) $\delta$ function\footnote{Inserting the more general $\prod_{x_E}\delta(W(\Delta\tau\Delta V_3\sqrt{g}-\Delta V_E))$, where $W$ is a generic function satisfying the regularity condition $W'(0)\neq0$, leads to an equivalent theory because just rescales the generating functional~(\ref{PIEucQG}) by a constant.}
\be \prod_{x_E}\delta(\Delta\tau\Delta V_3\sqrt{g(x_E)}-\Delta V_E), \label{deltaUG}\ee
where $\Delta V_E\equiv \Delta\tau\Delta V_3\omega_E$ is the fixed volume element at Euclidean spacetime point $x_E$, $\Delta \tau$ and $\Delta V_3$ are the imaginary time and the spatial volume elements (which become $d\tau$ and $d^3x$, respectively, in the zero lattice-spacing limit) and
$\omega_E$
corresponds to a  fixed non-dynamical volume form. The $\delta$ function in~(\ref{deltaUG}) can also be equivalently written as a functional integral:
\bea && \prod_{x_E}\delta(\Delta\tau\Delta V_3\sqrt{g(x_E)}-\Delta V_E) = \int \left(\prod_{x_E} \frac{dl(x_E)}{2\pi}\right) \nonumber \\ && \times\exp\left(i\int_{\tau_i}^{\tau_f} d\tau \int d^3x \, l(x_E)(\sqrt{g(x_E)}-\omega_E(x_E))\right), \nonumber\label{deltaexpUG}\eea
which corresponds to introducing an auxiliary field $l$ (a Lagrange multiplier).

The constraint factor~(\ref{deltaUG}) should also be inserted in the path integral~(\ref{PIEucQGFP}) for generic coordinate systems.  Note that $\Delta\tau\Delta V_3\sqrt{g}$ is an invariant volume element and, therefore, $\Delta\tau\Delta V_3\sqrt{g} = \Delta V_E$ is a physical (coordinate-independent) constraint. With this condition one maintains general covariance although the determinant of the metric $g$ is not dynamical~\cite{Weinberg:1988cp}. The insertion of~(\ref{deltaUG}) in~(\ref{PIEucQGFP}) thus leads to a physically distinct quantum theory, although, as we will discuss shortly, the classical limit is the same. Such insertion in particular implies that the operator corresponding to $g$ is reduced to a $c$-number function 
 in the unimodular theory. In~\cite{Salvio:2024joi} it was shown (without inserting~(\ref{deltaUG}) in the path integrand) that the Euclidean path integral of quadratic gravity is well defined in a physically acceptable region of the bare parameter space, solving the conformal-factor problem. Here we observe that the same constraints on the bare parameters still  ensure that the Euclidean path integral of unimodular quadratic gravity is well defined, i.e. even inserting~(\ref{deltaUG}), because~(\ref{deltaUG}) is a restriction on the functional integration domain.

When analytically continuing to real time,~(\ref{deltaUG}) gets replaced by the real-time version
\bea && \prod_{x}\delta(\Delta t\Delta V_3\sqrt{-g}-\Delta V)=\int \left(\prod_{x} \frac{dl(x)}{2\pi}\right) \times \nonumber \\  && \exp\left(i\int d^4x \, l(x)(\sqrt{-g(x)}-\omega(x))\right), \label{RdeltaUG}\eea
where now $\Delta t$ is the real-time element, $\Delta V\equiv \Delta t\Delta V_3\omega$ is the fixed volume element at Lorentzian spacetime point $x$ and 
$\omega$
corresponds to the  fixed non-dynamical volume form in the Lorentzian theory.  Analogously,~(\ref{RdeltaUG}) should  be inserted in the path integral~(\ref{PIEucGF2L}) for the generating functional of Green's functions\footnote{For a discussion of the path integral of Einstein gravity with the unimodular constraint see~\cite{Smolin:2009ti,deLeonArdon:2017qzg}.}.

One might doubt that quadratic gravity is still renormalizable after the unimodular constraint is imposed. To eliminate this doubt note that the constraint $\sqrt{-g}=\omega$ can be locally seen as a gauge fixing (the physical constraint is global, $\int d^4x\sqrt{-g}=\int d^4x\,\omega$). So quadratic gravity remains renormalizable because one can analyze loop diagrams using a gauge compatible with $\sqrt{-g}=\omega$; this is done, for example, in~\cite{Fiol:2008vk,deBrito:2021pmw}. Note that the proof of renormalizability of quadratic gravity in a generic gauge was provided in~\cite{Barvinsky:2017zlx}.

Suppose now that the action $S$ in~(\ref{PIEucGF2L}) instead of being only the classical action  also contains the effect of the matter fields that are functionally integrated out. Since the spacetime volume $\Delta t\Delta V_3\sqrt{-g}$ is non dynamical the vacuum-energy contribution of the matter fields, which can be absorbed in $\Lambda_0$, does not gravitate. This is an advantage of unimodular quadratic gravity: the CC is completely independent of the (too large) contribution coming from the known particles. One CC problem (the old one), which queries why the CC is not at least of the same order of the largest particle mass, is thus solved. Being the CC and the particle masses completely independent of each other, there is no reason why it should be.
This feature of unimodular quadratic gravity also allows us to non-perturbatively generate  the Planck scale through classically-scale invariant dynamics without a too large (Planckian) quantum-mechanically-generated CC~\cite{Donoghue:2018izj,Salvio:2020axm}.

If we now take the classical limit by following the methods of~\cite{Salvio:2024joi} (but with~(\ref{RdeltaUG}) present inside the path integral) we have to derive the field equations by imposing that $g$ is not dynamical, which leads to $g^{\mu\nu}\delta g_{\mu\nu}=0$, where $\delta g_{\mu\nu}$ is the variation of the metric that is performed in the stationary-action principle.
Nevertheless, in any UG (including unimodular quadratic gravity) this leads again to the same field equation one would have obtained without imposing $g^{\mu\nu}\delta g_{\mu\nu}=0$, although the CC emerges as an arbitrary integration constant rather than a coefficient in the action~\cite{Percacci:2017fsy}.  Therefore, the classical limit is the same. Note that from this argument it also follows that the physical CC is completely independent of $\Lambda_0$.

It is then important to understand whether the quantum difference between quadratic gravity and its unimodular counterpart could be observable.

  \section{Unimodular inflation}\label{Unimodular inflation}
  
  The natural arena to study quantum effects in gravity is inflation: cosmological perturbations emerge as quantum fluctuations in the theory of inflation. 

Let us then consider a cosmological spacetime.  Since $\omega$ does transform (like $\sqrt{-g}$) under general coordinate transformations with a well-known spacetime dependent factor, it is always possible to find a coordinate system where $\omega=a^4(u)$ where $a(u)$ is the cosmological scale factor and $u$ is the conformal time. This allows us to take a standard Friedmann-Lema\^itre-Robertson-Walker 
(FLRW) metric at the {\it classical} level: 
\be ds^2 = a(u)^2\left(\delta_{ij}dx^idx^j-du^2\right), 
\ee 
where we have neglected the spatial curvature parameter as during inflation the energy density is dominated by the scalar fields.
The possibility of taking the standard   FLRW metric reflects the fact that the unimodular condition enforced by~(\ref{RdeltaUG}) does not change the classical limit.

However, at quantum level the situation is different. This suggests that at linear order in the perturbations we may observe some differences because the perturbations are treated as quantum fields in the theory of inflation\footnote{If, on the other hand, perturbations are treated classically there is no hope to observationally distinguish between unimodular and non-unimodular gravity as the classical theory is the same.}.

The fact that in the formulation of unimodular gravity we are adopting general covariance is maintained allows us to use standard gauges. By  choosing the conformal Newtonian gauge, the metric describing the small perturbations around the FLRW spacetime can be written as 
\bea &&ds^2= a(u)^2\left\{ \left[(1-2\Psi(u, \vec{x})) \delta_{ij}+h_{ij}(u, \vec{x})\right]dx^idx^j \right. \nonumber \\ 
&& \left. + 2 V_i(u, \vec{x}) du dx^i  -(1+2\Phi(u, \vec{x})) du^2\right\}, \label{dsPert}\eea
where the vector perturbations $V_i$ satisfy
\be \partial_iV_i=0 \label{Cond1}\ee
 and the tensor perturbations $h_{ij}$ obey
 \be h_{ij}=h_{ji},  \qquad h_{ii} =0, \qquad \partial_ih_{ij}=0. \label{Cond2}\ee
Sometimes the Newtonian gauge is defined for the scalar perturbations $\Phi$ and $\Psi$ only  (see e.g.~\cite{Weinberg:2008zzc}). Here we consider a generalization, which also includes the non-scalar perturbations. A possible gauge-dependent divergence of $h_{ij}$ has been set to zero by appropriately choosing the gauge. 

Now, since $\Psi$, $\Phi$, $V_i$ and $h_{ij}$ are quantum fields, but the metric determinant $g$ is  reduced to a $c$-number function 
 in the unimodular theory, we must impose that any contribution to $g$ coming from these quantum fields  vanishes. In the conformal Newtonian gauge and at linear level in the perturbations \be g=-a^8(u)(1 + 2\Phi - 6\Psi),  \ee 
 where the traceless condition $h_{ii} = 0$ has been used, so we obtain the constraint
 \be \boxed{\Phi =3\Psi} \label{UGconstrI}\ee 
 in the unimodular theory.

Let us now assume for simplicity that inflation is driven by a minimally coupled scalar field, which happens to be a quasi-flat direction for the field values relevant during inflation. This can happen without fine-tuning if the inflaton is identified, for example,  with a pseudo-Nambu-Goldstone boson associated with an approximate and spontaneously broken global symmetry~\cite{Freese:1990rb,Salvio:2019wcp}. This type of inflation, known as natural inflation, is compatible with present cosmic microwave background (CMB) observations~\cite{Ade:2015lrj,Planck2018:inflation,BICEP:2021xfz} when implemented in quadratic gravity~\cite{Salvio:2022mld,Salvio:2021lka}. We can neglect the $R^2$ term in the action as the scalaron is assumed to be non-active during inflation in this setup.

The time-derivative of $\Phi$ does not appear in the action quadratic in the perturbations~\cite{Salvio:2017xul},
then $\Phi$  should be considered as a non-dynamical field. By varying  that action with respect to $\Phi$  one finds
\be -\frac{4}{3 f_2^2 \bp^2 a^2} \vec{\nabla}^4 \left(\Phi+\Psi\right) -6 {\cal H} \frac{d\Psi}{du}+2\vec{\nabla}^2 \Psi -6{\cal H}^2 \Phi = 0, \label{ConstraintPhi}\ee
where ${\cal H} \equiv \frac{1}{a}\frac{da}{du}$ and $\vec{\nabla}^4 \equiv (\vec{\nabla}^2)^2$ is the square of the spatial Laplacian $\vec{\nabla}^2$. Using now the unimodular constraint in~(\ref{UGconstrI}),
\be -\frac{16}{3 f_2^2 \bp^2 a^2} \vec{\nabla}^4\Psi -6 {\cal H} \frac{d\Psi}{du}+2\vec{\nabla}^2 \Psi -18{\cal H}^2 \Psi = 0. \label{ConstraintPhiUG}\ee
By performing a Fourier transform on the spatial coordinate, 
\be  \Psi(u,\vec{x}) =  \int \frac{d^3q}{(2\pi)^{3/2}}e^{i \vec{q}\cdot\vec{x}}  \tilde\Psi(u,\vec{q}) \ee
this equation reads
\be -\frac{16 q^4}{3 f_2^2 \bp^2 a^2}\tilde\Psi -6 {\cal H} \frac{d\tilde\Psi}{du}-2 q^2 \tilde\Psi -18{\cal H}^2 \tilde\Psi = 0, \label{ConstraintPhiUGF}\ee
where $q\equiv |\vec q|$. Using the de Sitter expression $a(u)=-1/(Hu)$, where $H$ is the inflationary Hubble rate, one finds that the general solution of~(\ref{ConstraintPhiUGF}) is 
\be \tilde\Psi(u,\vec{q}) = \exp(q^2u^2/6+2 H^2q^4u^4/(9f_2^2\bp^2))\,u^3 \,{\cal C}\ee 
with ${\cal C}$ a generic operator that is constant in $u$. 

The main phenomenologically interesting regime is the superhorizon limit, $u\rightarrow 0^-$, when $a\rightarrow+\infty$. In this limit $\Psi\to0$ as fast as $u^3$. It is then possible to show that the standard curvature perturbation ${\cal R}$ acquires the expression in Einstein gravity~\cite{Salvio:2017xul}. The predictions for the tensor-to-scalar ratio $r$ and the spectral index $n_s$ is then the same as in quadratic gravity without the unimodular constraint. However, the fact that $\Psi\to0$ in the superhorizon limit also implies that the extra isocurvature mode $B$ present in quadratic gravity, as shown in~\cite{Salvio:2017xul,Salvio:2020axm}, decouples in unimodular quadratic gravity. Since future CMB observations may detect the power spectrum of $B$~\cite{Salvio:2019ewf}, we conclude that quadratic gravity can be distinguished from its unimodular counterpart: the former predicts an isocurvature mode that is absent in the latter.

  \vspace{-0.1cm}
   
\section{Post-inflationary cosmology}\label{Post-inflationary cosmology}

After inflation a period of reheating should take place. In order not to introduce a large fine-tuning of the Higgs mass the inflaton should belong to a somewhat hidden sector. Reheating can take place, for example, thanks to the presence of several light and weakly coupled scalar fields, which have sizable couplings to the observed particles~\cite{Salvio:2019wcp}. This situation is typical in asymptotically free Standard Model extensions~\cite{Giudice:2014tma,Pelaggi:2015kna}. The aforementioned  scalar fields undergo quantum fluctuations that are of order $H/(2\pi)$ independently of the presence of the unimodular constraint: those fluctuations emerge as solutions of the linearized equations of those scalar fields on the inflationary de Sitter background and such equations are independent of the unimodular constraint. This mechanism ensures that the energy density of the inflaton is transferred (as radiation) to the observable sector, which includes the Standard Model (SM) fields at low energy. 

In both the inflationary and subsequent radiation-dominated epochs life is clearly impossible. Indeed, in the inflationary epoch the matter density is effectively absent and the anthropic bound of~\cite{Weinberg:1987dv,Weinberg:1988cp} is not satisfied; in the  radiation-dominated epoch the universe is too hot. As time passes by the radiation energy density decreases and the temperature drops so that a matter-dominated universe emerges at some point, as the SM features more massive than massless degrees of freedom. Since the matter density $\rho_M$ also decreases with time, eventually the energy density due to the CC, $\rho_\Lambda$, overcome $\rho_M$ again. In order not to violate the anthropic bound of~\cite{Weinberg:1987dv,Weinberg:1988cp}, $\rho_\Lambda$ should not be much larger than $\rho_M$. Since life takes time to develop, it is reasonable to find a scientific community able to measure the CC  at the latest possible epoch compatible with this bound, which is when we live.
Note that the value of the CC is here explained\footnote{This   also explains why $\rho_M$ and $\rho_\Lambda$ have the same order of magnitude today (the coincidence problem).} anthropically with a multiverse made by different eras in a single big bang; this type of multiverse was mentioned before, see e.g.~\cite{Banks:1984cw,Weinberg:2005fh}. 

Note that, since unimodular gravity solves the old issue of explaining why the CC is not at least as large as the largest particle mass scales, this multiverse (multiple universes across time) does not need to feature a complex landscape for the CC, unlike in non-unimodular (standard) gravity.

  \vspace{-0.1cm}
   
\section{Conclusions}\label{Conclusions}

  \vspace{-0.2cm}
  
  Here, an unimodular version of a non-perturbative and background independent quantum gravity featuring quadratic-in-curvature terms has been constructed and the cosmological constant problems have been addressed. General covariance is preserved.
  
  It was shown that the unimodular condition affects the quantum predictions of the theory; in particular an isocurvature mode, which is within the reach of future CMB observations, is removed by unimodularity. This allows us to physically distinguish between standard and unimodular gravity, although the two theories share the same classical limit. 
  
  Although unimodular gravity explains why the CC is not as large as the largest particle mass scale (the old CC problem), because the CC is completely independent of the vacuum energy, the new CC problem (why the dark energy and matter densities are comparable?) calls for other ingredients. To address this further issue a multiverse made by different eras in a single big bang was considered and the observed value of dark energy is explained anthropically, but without the need of a huge landscape: the dark energy density is not constant, but varies during the various eras, such that in the period with the largest probability of hosting intelligent life the dark energy density is larger than (but of the same order of magnitude as) the matter density.

\vspace{-0.2cm}

\section*{Acknowledgments} 
\vspace{-0.1cm} 

\noindent    This work was partially supported by the Italian Ministry of University and Research (MUR) under the grant PNRR-M4C2-I1.1-PRIN 2022-PE2 Non-perturbative aspects of fundamental interactions, in the Standard Model and beyond F53D23001480006 funded by E.U. - NextGenerationEU.

\vspace{0.cm}



\end{document}